\documentclass[sigconf]{acmart} 

\usepackage[utf8]{inputenc}        
\usepackage{csquotes}              
\usepackage[nameinlink]{cleveref}  
\usepackage[subtle]{savetrees}

\AtBeginDocument{%
  \providecommand\BibTeX{{%
    \normalfont B\kern-0.5em{\scshape i\kern-0.25em b}\kern-0.8em\TeX}}}

\copyrightyear{2022} 
\acmYear{2022} 
\setcopyright{rightsretained}
\acmConference[ITiCSE 2022] {Proceedings of the 27th ACM Conference on Innovation and Technology in Computer Science Education Vol 1}{July 8--13, 2022}{Dublin, Ireland.} \acmBooktitle{Proceedings of the 27th ACM Conference on Innovation and Technology in Computer Science Education Vol 1 (ITiCSE 2022), July 8--13, 2022, Dublin, Ireland}
\acmPrice{15.00}
\acmDOI{10.1145/3502718.3524819}
\acmISBN{978-1-4503-9201-3/22/07}

\begin{document}

\title{Process Mining Analysis of Puzzle-Based Cybersecurity Training}


\author{Martin Macak}
\affiliation{%
  \institution{Masaryk University}
  \streetaddress{Botanicka 68a}
  \city{Brno}
  \country{Czech Republic}}
\email{macak@mail.muni.cz}
\orcid{0000-0001-9655-9228}

\author{Radek Oslejsek}
\affiliation{%
  \institution{Masaryk University}
  \streetaddress{Botanicka 68a}
  \city{Brno}
  \country{Czech Republic}}
\email{oslejsek@mail.muni.cz}
\orcid{0000-0002-0562-6892}

\author{Barbora Buhnova}
\affiliation{%
  \institution{Masaryk University}
  \streetaddress{Botanicka 68a}
  \city{Brno}
  \country{Czech Republic}}
\email{buhnova@mail.muni.cz}
\orcid{0000-0003-4205-101X}


\fancyhead{}

\begin{abstract}
The hands-on cybersecurity training quality is crucial to mitigate cyber threats and attacks effectively. However, practical cybersecurity training is strongly process-oriented, making the post-training analysis very difficult. This paper presents process-mining methods applied to the learning analytics workflow. We introduce a unified approach to reconstruct behavioral graphs from sparse event logs of cyber ranges. Furthermore, we discuss significant data features that affect their practical usability for educational process mining. Based on that, methods of dealing with the complexity of process graphs are presented, taking advantage of the puzzle-based gamification of in-class training sessions.
\end{abstract}



\begin{teaserfigure}
  \includegraphics[width=\textwidth]{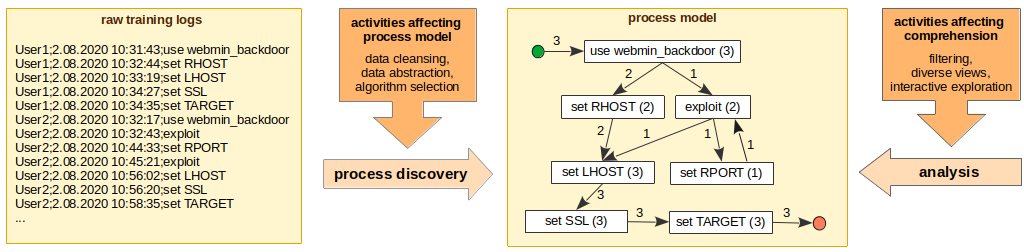}
  \caption{Principles of process discovery. Although the event log (left) captures a simple use of the Metasploit tool by only three users, it is very difficult to analyze. Process models (right) provide better cognitive features simplifying comprehension. However, the practical usability depends on many factors affecting the process discovery and process model analysis.}
  \label{fig:teaser}
\end{teaserfigure}

\maketitle

\section{Introduction}
\label{s:introduction}

As cybersecurity skills require higher-order thinking, the best way to develop and ameliorate these abilities is through hands-on training~\cite{mcmurtrey2008,hatzivasilis2020}. 

In many learning areas, hands-on training produces a tangible output, e.g., a code that can be checked, analyzed, and evaluated. However, this is not the case of cybersecurity, where learning goals consist of process-oriented tasks related to attack or defense skills, e.g., scanning the network for vulnerable servers or protecting the server by a firewall. Modern cyber ranges, in which practical cybersecurity training is often organized~\cite{Knupfer2020,ukwandu2020,yamin2020,chouliaras2021}, provide only limited data about these tasks in the form of event logs, which makes it hard to answer questions like ``Where did the trainees get in trouble and why?'' without deep insight into their behavior. 

Process mining (PM) methods have great potential in answering analytical questions in process-oriented application domains~\cite{PMbook}. Our goal is to utilize them for cybersecurity training analysis. 
However, the quality and practical usability of models produced by existing PM algorithms are influenced by many factors, as schematically depicted in Figure~\ref{fig:teaser}. This paper addresses open questions that affect the usability of existing PM techniques for scalable post-training analysis of hands-on cybersecurity exercises, aiming to provide guidelines for their usage. Specifically, we formulated the following research questions.

\textbf{RQ1: Data Abstraction -- how to convert cyber training data to the format suitable for process mining?} The PM workflow is used to find a descriptive model of the process. The idea lies in transferring event logs captured during training sessions into behavioral graphs that would provide better cognitive features for understanding users' behavior than the raw data. As the data collected from hands-on training environments are highly variable, the suitable mapping to the input of PM algorithms is unclear. Our goal is to propose a unified data abstraction that can serve as a generic approach to data preprocessing in different cyber ranges with various data sources.

\textbf{RQ2: Process Discovery -- what are the key obstacles in the process discovery phase and how to overcome them?} Having the data in the format suitable for process discovery algorithms does not guarantee that generated process models are reasonable. Many input data characteristics affect the process discovery and its practical usability for educational process mining. We aim to identify key features of cyber training data and discuss the limits and obstacles of the process discovery phase.

\textbf{RQ3: Exploratory Analysis -- how to deal with the complexity of process graphs during analysis?} Even a relevant process graph obtained by process discovery can become too big and complex for practical learning analytics. Their cognitive features can decrease and become similar to searching for answers in the raw data. Therefore, we research exploratory tactics that can help to tackle the graph complexity problem. We combine specific features of the cybersecurity training data with well-established visual analysis approaches to discuss and demonstrate their usability for interactive learning analytics.

In this paper, we restrict ourselves to the in-class (i.e., supervised) \emph{Capture The Flag} (CTF) cybersecurity games~\cite{werther2011,vigna2014,davis2014,svabensky2018,oslejsek2019}. They follow puzzle-based gamification principles, where puzzles are used as a metaphor for getting students to think about how to frame and solve unstructured problems~\cite{michalewicz2008}. This training type is very popular in the education of beginners, e.g., students of cybersecurity courses.

\section{Related Work}

The idea of using process mining methods for learning analytics is not new. The term educational process mining (EPM) is often used for the application of PM to raw educational data~\cite{romero2016}. Bogarin et al.~\cite{RWBogarin2018education} provide a comprehensive overview of EPM techniques and challenges. We use their classification of event log challenges to clarify features of cybersecurity CTF games and then to address key hurdles, especially the data complexity.

PM has already been applied in numerous specific educational situations. Macak et al.~\cite{macak2021using} used process mining to understand student coding-behavior patterns from the Git log of their projects. Mukala et al.~\cite{rwMukala} use process mining to obtain and analyze students' learning habits from Coursera. It is also used to detect students' learning difficulties by Bannert et al.~\cite{rwBannert}. Multiple other approaches use process mining to gain a better understanding of the educational process from Moodle logs~\cite{rwBogar,romero2016}. Our work shares some ideas and principles with these approaches. Still, it addresses a different application domain --- puzzle-based cybersecurity training, aiming to utilize the specific data properties to deal with process mining challenges.

Some papers also directly address the utilization of process mining for the analysis of hands-on cybersecurity exercises~\cite{weiss2016,weiss2017,andreolini2020,svabensky2021}. These approaches demonstrate that directed graphs carefully constructed from command history can provide useful information about trainees' behavior. The graphs are built on restricted data samples to deal with the complexity --- only commands identified as significant are selected. Our research is more generic, covering a wider variety of training events and enabling to extent process modeling with new data types. Another difference is in the conceptual approach. The previous approaches are based on conformance checking, i.e., monitoring and checking whether the real execution of the process corresponds to the reference process model. Our solution focuses on exploratory learning analytics based on general process discovery methods where no process model of expected behavior is assumed.

Mirkovic et al.~\cite{mirkovic2020} use terminal histories and exercise milestones to enable automated assessment of hands-on cybersecurity training. 
In contrast, our work aims to introduce a human into the analytical loop~\cite{oslejsek2018} when process models are reconstructed from logs automatically but then interactively analyzed by domain experts so that they are able to reveal hidden relationships in the data.

\section{RQ1: Data Abstraction} \label{sec:data-abstraction}

In this section, we classify cybersecurity training data and provide their unified mapping into the input of process discovery algorithms to address the research question \emph{RQ1}.
The proposed solution builds on our long-term experience in developing \emph{KYPO Cyber Range}\footnote{KYPO is a Czech acronym for Cybersecurity Polygon.}~\cite{vykopal2017}.
that we are operating since 2013 and which serves as a platform for regular practical training of students of our university. We also closely collaborate with domain experts (cybersecurity educators) from our university who represent target users of post-training analysis tools. 

\subsection{Principles of CFT games}

Cybersecurity CTF games consist of well-described cybersecurity goals divided into consecutive tasks (puzzles). Completing each task yields a text string called the \emph{flag} that must be inserted into the system to proceed to the subsequent task. Moreover, trainees can take hints or skip the entire task. Points are awarded or deducted for these actions so that the final scores of individual trainees are mutually comparable and can be used for their basic evaluation. 

Trainees perform cybersecurity tasks on remote hosts located inside isolated computer networks. Modern cyber ranges provide a virtualized implementation of such networks, where each trainee has its own copy of the network and is able to access hosts via remote command lines or desktops, likewise in the physical world.

\subsection{Data Types} \label{sec:data-description}

Cyber ranges provide telemetry data in the form of events. They are captured from multiple sources and can provide different levels of granularity~\cite{svabensky2021monitoring}. This paper discusses three distinct data categories, but other data types or levels can be included if available.

\begin{table}[t]
\centering
\caption{Game events and their meaning.}\label{tab:gameEvents}
\begin{tabular}{ |c|l|  }
\hline
\textbf{Event} & \textbf{Description: The trainee ...}  \\
\hline
TrainingRunStarted & \dots started the training.\\ \hline
TaskCompleted & \dots submitting a correct flag. \\ \hline
WrongFlagSubmitted &  \dots submitted a wrong flag. \\ \hline
HintTaken & \dots took a hint. \\ \hline
SolutionDisplayed & \dots viewed the task solution. \\ \hline
\end{tabular}
\end{table}

\textbf{Game events} are produced by the gaming interface of KYPO Cyber Range. Its goal is to provide instructions and guide trainees through the whole training session. User interaction with the interface produces events that capture the gameplay state in the training scenario. Events summarized in Table~\ref{tab:gameEvents} reflect the puzzle-based principles, and they are typical for all games regardless of the content. 

\textbf{Bash commands} are produced inside computer networks in which the cybersecurity tasks are solved. Currently, commands executed on the UNIX command line (shell) are available.

\textbf{Metasploit tool} capture the usage of the Metasploit framework -- a popular command-line application used for penetration testing. These events are also captured at hosts of the computer network, likewise the bash history, but represent even a finer-grained type of data.

Each event, regardless of its type or granularity, is extended with additional pieces of information, such as the trainee's identifier, timestamp, and the task (puzzle) in which the event appeared. Moreover, individual event types can have specific mandatory or optional data. For example, the submission of the flag always includes also the flag value, and commands may include their parameters.

\subsection{Unified Data Mapping}

All process mining techniques require the presence of data with specific semantics~\cite{PMbook}: (a) Each event in the log file needs to refer to a single process instance, named \emph{case}, (b) each event needs to refer to a step in the process, named \emph{activity}, and (c) events within the case have to be ordered, either sequentially or by including \emph{timestamp}.
These minimal requirements ensure that each case represented by the sequence of activities can be treated as a \emph{trace} of user actions, enabling the process discovery algorithms to produce graph models capturing all the traces compactly.

Unfortunately, CFT data are highly variable. They can differ in abstraction and semantics, as shown by the three aforementioned data types --- game events, bash commands, and the Metasploit tool. 

To deal with variability, we introduce a generic data abstraction layer that makes the mapping smooth and transparent regardless of the specific training content. The proposed classification scheme serves as a mediator between heterogeneous event logs of the cyber range and the data format required by process mining techniques. Table~\ref{tab:abstraction} provides a mapping example, where a snippet of CTF data (four events) is mapped into the abstraction layer.

\begin{table*}[htb]
\centering
\caption{Mapping of raw data onto the unified CTF data abstraction and process mining inputs.}\label{tab:abstraction}
\begin{tabular}{|l|c|c|c|c|c|}
\hline
\textbf{Process mining input:} &     & \emph{activity}  &                    & \emph{time/ordering}  & \emph{caseID}  \\ \hline
\textbf{Data abstraction: }  & \textbf{\textsc{event type}} & \textbf{\textsc{event}} & \textbf{\textsc{event parameters}} & \textbf{\textsc{timestamp}} & \textbf{\textsc{trainee}} \\\hline
\textbf{Raw data (snippet): }    & game       & HintTaken 41-1 &                    & 2020-05-14 10:16:11 & user 1  \\ 
                                 & game       & HintTaken 41-2 &                    & 2020-05-14 10:16:34 & user 1  \\ 
                                 & msf        & exploit        & -j                 & 2020-05-14 10:18:23 & user 2  \\ 
                                 & bash       & nmap           & -sL 10.1.26.9:5050 & 2020-05-14 10:32:16 & user 1  \\ 
\hline
\end{tabular}
\end{table*}

\textbf{\textsc{Event type:}} A rough classification defining different types of events. We can distinguish player actions in the game from reactions of the system on the player's actions, or assessment events, for instance~\cite{salen2004}. The exact classification used for process mining depends on available data and analytical goals. Usually, each event type has its specific structure (required or optional data), and they affect how the pieces of information are spread across the other elements of the data abstraction. This paper deals with three aforementioned event types: \texttt{game} events capturing the players' progress, \texttt{bash} commands used on network hosts, and \texttt{msf} for Metasploit commands also used on hosts. Additional types can be easily defined. 

\textbf{\textsc{Event:}} Finer classification of \textsc{event types}. Events represent a primary subject of behavioral analysis. They should capture significant steps in the development of training sessions. Therefore, they represent the \emph{activities} of the process models. In CTF games, events are either game events defined in Table~\ref{tab:gameEvents} or any shell or Metasploit commands.

\textbf{\textsc{Event parameters:}} Optional data associated with \textsc{events}. Additional information that extend \textsc{events} and enable the analyst to distinguish finely between them. For example, shell or Metasploit commands can have additional arguments, or the \emph{HintTaken} game event can be equipped with a short hint description. 

\textbf{\textsc{Timestamp:}} A timestamp of the \textsc{event}. This is required because of the aggregation of multiple \textsc{event types}.

\textbf{\textsc{Trainee:}} An anonymized unique identifier of the trainee who produced the \textsc{event}. Using the \textsc{trainee} identifier as the \emph{caseID} ensures that the process discovery reconstructs the walkthroughs of individual trainees. The walkthrough perspective presents a primary subject of learning analytics. It enables analysts to compare trainees' behavior mutually as well as to analyze the expected versus anomalous behavior with respect to the training definition. 

\begin{figure}
    \centering
    \includegraphics[scale=0.45]{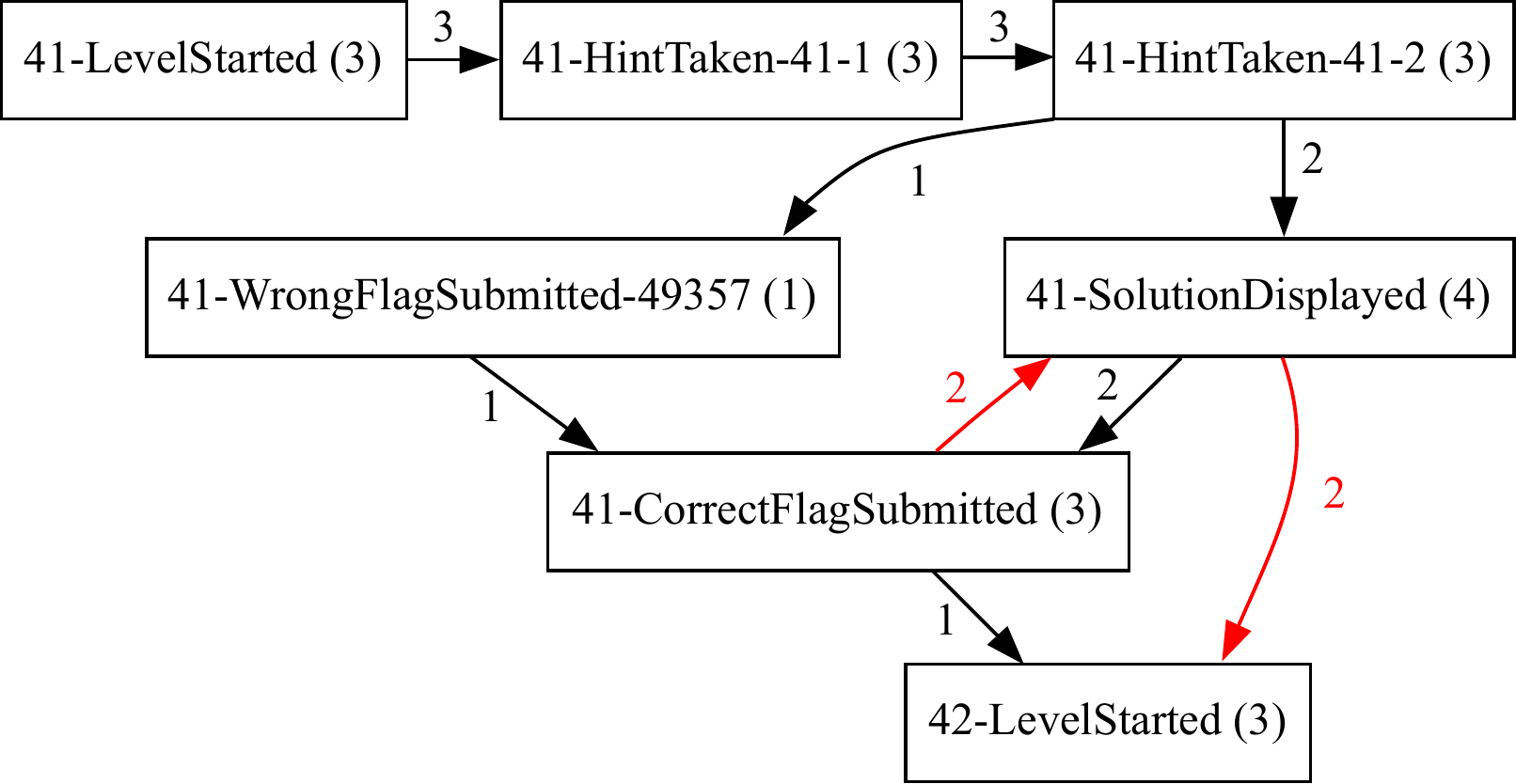}
    \caption{Level of detail based on event type, shown on task~41. The red arrows highlight a flow in the gameplay caused by a bug in platform implementation.}
    \label{fig:flow}
\end{figure}

\subsection{Usability}

We conducted practical experiments with PM4Py~\cite{pm4py} library to check the utilization of CTF data mapping for process discovery. The experiments proved that the proposed workflow is able to generate meaningful process models from our raw data at runtime. 

Our initial research revealed that heuristic nets especially fit the analytical goals of educators the best. Even though the goal of these experiments was not to study the impact of obtained models on learning analytics, we were able to notice several interesting facts from process graphs. For example, the game task captured in Figure~\ref{fig:flow} can be considered tough for the trainees because they all took two hints, then two gave up (they looped at the solution with the correct flag). One trainee found and submitted the correct flag on the second try. 
Using the same model, we also have discovered a flaw in the implementation of KYPO Cyber Range. If trainees submitted the correct flag after displaying a task solution, they were redirected to the solution page instead of being moved to the next puzzle.  


\section{RQ2: Process Discovery} \label{s:rq2}

The proposed data abstraction enables us to use cyber-training data transparently in process mining algorithms. However, the practical usability of process discovery for learning analytics is affected by many factors. 

The most significant challenges that appear when using event logs for educational process mining are addressed in the previous research~\cite{RWBogarin2018education,romero2016}. Using their classification, we analyzed raw data of multiple training sessions to identify the most significant features that affect the utilization of cyber training logs in post-training process discovery. Fifteen training sessions of six different CTF games were analyzed for statistical properties like training duration or the number of log events of different types. 

In what follows, we summarize our observations and lessons learned. Each data characteristic is introduced by the problem statement followed by our findings.

\subsection{Data Size}

The number of cases or events in event logs may become so high that they exceed the time or memory requirements of process discovery algorithms. Moreover, as we aim at providing interactive data exploration, the speed of PM generation should be close to real-time.

The real amount of collected data depends on the number of participants, the difficulty of the training content (i.e., the amount of potentially recorded activities), and training duration. CTF games are intended primarily for beginners and then relatively small -- consisting of 4-6 cybersecurity tasks (puzzles) solvable in roughly 120 minutes (observed minimum was 65, maximum 210, average 119). Moreover, in-class training sessions considered in this paper restrict the data even more. It is because also the number of trainees is limited. In our datasets, the number of trainees varied between 4 and 20, 10 on average.

The number of logged events (and then potential nodes of the process graphs) in our datasets varied between 370--3000 per the whole training session (average 1100, median 814) and between 53--150 per participant (average 108, median 111). 


The data amount does not pose any problems to current process discovery algorithms that can treat such an amount of data very quickly~\cite{Hernandez2015}. 
On the other hand, our experiments revealed that the
problem could be with the comprehensibility of produced models, even for limited in-class CTF data. Obtained graphs consisting of tens or hundreds of nodes are usually too complex to be cognitively treated by analysts. Employment of data filtering and interactive exploration is, therefore, necessary.

\subsection{Distribution of Data Sources}

Data for educational process mining may be distributed over a variety of sources, e.g., theory and practice classroom or online learning environments. These sources provide event logs with different structures and meanings, which makes their unification and aggregation for process discovery challenging.

We focus on only the data collected from cyber ranges. Considering other supporting sources of information is out of the scope of this paper. However, the problem with data distribution occurs as well because modern cyber ranges produce data from at least two data sources: (a) the gaming environment responsible for the training tasks and learning milestones and (b) sandboxes (computer networks) where the tasks are performed. Nevertheless, the data abstraction discussed in Section~\ref{sec:data-abstraction} solves this problem by unifying the way the data from multiple data sources is handled and used transparently for process discovery.

\subsection{Granularity}

Events in event logs may have a different level of detail. The data presented in Section~\ref{sec:data-abstraction} demonstrates that this problem also occurs in the cyber training logs. 

Thanks to the unified data abstraction, the granularity poses no technical problem for the process discovery of cyber training logs because the data can be mapped transparently. Moreover, this unified approach brings significant advantages when data with different granularity are put together. Multiple levels of granularity can define multiple levels of abstraction that can drive data exploration strategies, as discussed in Section~\ref{s:rq3}. 

\subsection{Noise in the Data}

Noise is defined as exceptional behavior which is not representative of the typical behavior of the process.

We observed that many students leave the training unfinished due to the lack of time or the loss of motivation. Normally, this kind of noise could pose troubles for educational PM analysis. However, the puzzle-based structure of CTF games provides clear milestones --- correct flags that explicitly delimit borders of training phases. Therefore, it is easy to spot this situation in process graphs and further investigate the reason. In general, the puzzle-based structure can be considered a template of expected behavior, and any difference revealed by PM models can indicate flows in the game design or the training organization worth further investigation.

\subsection{Incompleteness}

Possible data incompleteness is tightly connected to the measurement infrastructure and techniques. Although the individual cyber ranges can differ in these aspects, they are often complex, distributed with asynchronous computation, based on underlying virtualization, and then unreliable. Our long-term experience with organizing cyber training sessions of many types reveals that many things can go wrong due to failures in low-level virtualization services, network connectivity, or improper usage. These failures then cause missing data.

The experiments have shown that keeping this incomplete data in the dataset can produce biased process models. Unfortunately, it is usually very difficult to notice from process graphs that there is something wrong with the raw event logs. Data cleansing and completeness checking have to be usually done in the preprocessing phase of the analytical workflow.

\subsection{Timestamps}\label{s:timestamps}

Events need to be ordered per case to be used for process discovery. Our experiments revealed three significant issues in the collected cyber training data.

Distributed environments like cyber ranges can produce timestamps that are not sufficiently synchronized. Only a small shift in times can re-order events and then produce significantly different process models. Therefore, precise synchronization at the level of the underlying infrastructure is required.

Trainees can start the exercise at different times, even if they sit in the same classroom. This aspect becomes even more significant if the training is not organized as a fixed-time group session, but the training content is available online at any time. Fortunately, the puzzle-based structure of CTF games enables us to identify the exact start of the gameplay of individual trainees and then compute relative
times instead of using absolute times, obtaining meaningful
process models.

An even worse situation can appear if the trainees can ``pause the training''. It does not happen on session-based training courses with a tight schedule. However, loosely conceived training programs would enable participants to stop playing for a while and continue with tasks later, even the next day. 
Therefore, datasets from the loosely organized training events require much more attention and expertise to be paid by the analyst, who has to take care of time corrections and interpretation of obtained models.



\section{RQ3: Exploratory Analysis} \label{s:rq3}

Despite the limited size of event logs produced by in-class CTF games, the experiments turned out that obtained process graphs can be too complex and incomprehensible for effective analysis. As the complexity of process graphs can pose a critical aspect for practical usability, we discuss possible strategies for tackling this problem in this section.

\subsection{Filtering Driven by Data Abstraction}

The granularity of training logs discussed in Section~\ref{s:rq2} can be used to control the level of detail and then the size of obtained process graphs. This kind of semantic classification can be used for efficient data filtering and implementing the well-known Shneiderman's visual information-seeking principle: Overview first, zoom and filter, then details-on-demand~\cite{shneiderman1996}.

The granularity is encoded in the \emph{Event types} data abstraction parameter that defines different semantic views of the data. CTF \emph{game events} delimit boundaries of individual puzzles in which other events appear. On the other hand, \emph{bash commands} represent a detailed view of solving tasks within a puzzle. \emph{Metasploit commands} also provide a similar view but at an even more fine-grained level of detail -- the usage of a specific hacking tool. 
Based on this observation, filtering of the process model to only a specific \textsc{event type} could provide the desired level of detail and then reduce information complexity. Figure~\ref{fig:teaser} depicts the model limited to the Metasploit only, while in Figure~\ref{fig:flow}, only game events are selected, entirely omitting bash and Metasploit commands.

Another graph complexity reduction technique utilizes the distribution of the raw data between \emph{event} and \emph{event parameter}. Consider the situation when a trainee takes a hint 41-1 and then a hint 41-2. If these events are mapped into the data abstraction like in Table~\ref{tab:abstraction}, then the process discovery algorithm distinguishes between hints 41-1 and 41-2, creates separate nodes for them and produces a model like that in Figure~\ref{fig:flow}. On the contrary, if we change the mapping so that \textsc{event} = ``\emph{HintTaken}'' and hint numbers 41-1 and 41-2 are provided only as \textsc{event parameters}, then a simplified model is produced with only a single joint ``41-HintTaken'' node covering all hints taken in the task.

This filtering principle is even more important for Bash and Metasploit commands than game events because trainees have big freedom of what to type on the command line. Mapping only command names without parameters to \textsc{events} seems to be a reasonable strategy for initial analysis. On the other hand, an \texttt{ssh} command, for instance, says nothing about the remote connection that has been made. In this case, the analyst should rather map the connection argument to the \textsc{event} and then produce separate nodes like \texttt{``ssh root@172.18.1.5''} and \texttt{``ssh admin@172.18.1.5''} in the process graph. Only then, the analyst is able to explicitly see different attempts (traces) and evaluate their correctness. 
Therefore, the mapping has to be used carefully and iteratively during the analytical process to balance information hiding with graph complexity. 

\subsection{Puzzle-based Fragmentation and Drill Down}

\begin{figure}
    \centering
    \includegraphics[scale=0.22]{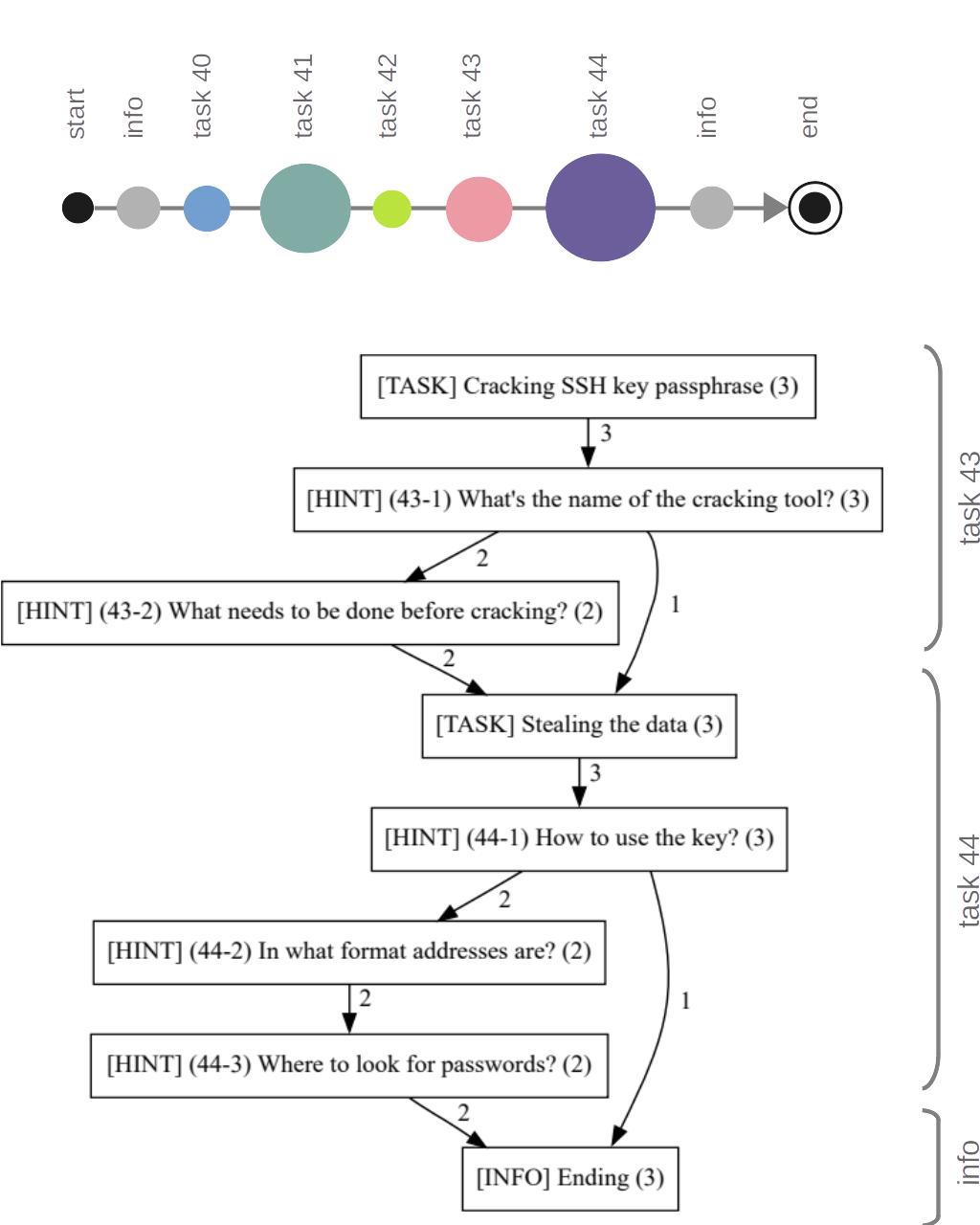}
    \caption{Approximate visualization of tasks difficulty (up) and corresponding process graph (bottom). Only \emph{game events} are included in the process graph. Only tasks 43, 44, and the info puzzle are shown in this example to save space.}
    \label{fig:fragmentation}
\end{figure}

While the unified data abstraction can serve as a fine-grained filtering mechanism of the entire process model across multiple puzzles, the puzzle-based structure of the training content provides a vertical fragmentation of the data usable for drill-down exploration. Process graphs can be logically split into loosely coupled coherent parts that correspond to individual puzzles, as shown in Figure~\ref{fig:fragmentation}, where the puzzles of tasks 43, 44, and an info puzzle are visually recognizable.

Based on this observation, we can tackle the complexity of multi-puzzle graphs by allocating puzzles' boundaries and using obtained graph fragments for coarse-grained filtering and data aggregation. Statistical data of each fragment (puzzle) can be distilled into an overview of the whole training session, while smaller detail process graphs of individual puzzles can be used for detail-on-demand exploration. 

Figure~\ref{fig:fragmentation} illustrates this multi-layer approach. A high-level overview of the training session consists of a series of circles whose size and color can encode interesting metrics. For example, spheres' size can be calculated from the number of activities -- nodes of the puzzle's process graph. In this case, the size of task 44 would indicate that it is the most complicated part of the training with a lot of recorded activities (note that Bash and Metasploit commands are omitted from the graph view in Figure~\ref{fig:fragmentation}, but they can be calculated in the complexity metric and reflected in the sphere's size). Then the analyst can interactively drill down into selected puzzles to analyze the reason that could be either the task complexity (it requires many commands to be used) or difficulty (trainees struggled with the task completion and then generated many events).

The number of activities discussed in the previous example is not the only possible metric. Alternate metrics can provide a different perspective on the training results. For example, the number of displayed solutions can indicate how many trainees gave up the puzzle due to the task's difficulty, demotivation, or lack of time.


\section{Discussion and Future Work}

This section summarizes our results, putting emphasis on simplifications that we put on the training data.

\subsection{Limitations}

Results of this study are limited by two key constraints put on hands-on training: puzzle-based gamification and in-class education.

Considering only well-structured puzzle-based CTF games enables us to better classify data and fragment complex process models for drill-down exploration. We omit other training concepts, e.g., complex Cyber Defense Exercises~\cite{patriciu2009} that are intended for experienced professionals. They use wide network topologies, provide freedom in the exercise scenarios to simulate reality, and participants collaborate in teams, which may produce higher amounts of less-structured data and then violate prerequisites of our observations. 

In-class training is limited to time and number of participants, which helps us keep data size within reasonable limits. However, cybersecurity training programs can also have the form of online courses accessible at any time by an unlimited number of participants. These courses can produce a significantly larger amount of data and pose some troubles with time dependencies that are crucial for correct process discovery, as discussed in Section~\ref{s:timestamps}. Therefore, extending our approach beyond in-class teaching requires further research.

We are also aware that data sets used in this paper were collected from CTF games organized by a single team in a single cyber range, which could affect our observations and limit generalization. On the other hand, to the best of our knowledge, other modern cyber ranges supporting CTF training style, e.g., Cyris, CyTrONE, or Ares~\cite{pham2016,beuran2018, Ares}, share the same concepts and principles that are discussed in this paper. Therefore, the training content and collected data may differ in detail, but the key aspects like size or types of events are very similar.

\subsection{Implications for Teaching Practice}

It is tough to identify flows in training design or analyze trainees' behavior without transforming events into models with better cognitive features. The proposed unified data abstraction can be directly used to map the data from cyber ranges into the input of exiting process mining tools and algorithms. 

On the other hand, the practical usability of these generic tools can be limited. Their usability depends on the support of discussed data filtering and exploration techniques that are often domain-specific. Therefore, we are currently working on integrating these techniques into the analytical interface of KYPO Cyber Range so that the process mining analysis becomes an integral part of the training life cycle.

\section{Conclusion}

This paper explores the practical usability of existing process mining algorithms to analyze cybersecurity training sessions and provides observations that support this direction. 

Despite the variability of data collected from cyber ranges, we proposed a unified data abstraction for using the data as the input required by process mining algorithms. We tested the usability with data captured in a cyber range that we operate. The practical experiments proved the usefulness of our approach for answering questions related to learning analytics and evaluating corresponding hypotheses but also revealed limits caused by concrete features of the raw data.

We analyzed data from 15 training sessions to reveal significant features that affect the practical usability of process discovery algorithms. The main problem we faced was the complexity of the obtained graphs. Therefore, we introduced several strategies of data filtering and interactive data exploration that are built on the features of puzzle-based in-class form of exercises.

\begin{acks}
This research was supported by the Security Research Programme of the Czech Republic 2015–2022 (BV III/1–VS) granted by the Ministry of the Interior of the Czech Republic under No. VI20202022158 – Research of New Technologies to Increase the Capabilities of Cybersecurity Experts.
\end{acks}

\bibliographystyle{ACM-Reference-Format}
\bibliography{bibliography}

\end{document}